\documentclass[twocolumn,aps,prl]{revtex4-2}

\usepackage{amsmath, amssymb, physics, bm}
\usepackage{graphicx}
\usepackage{tikz}

\usepackage[T1]{fontenc}

\usepackage{floatrow}
\usepackage[caption=false]{subfig}

\DeclareMathOperator{\UConf}{UConf}
\DeclareMathOperator{\Hom}{Hom}

\begin{document}
\title{Eigenvalue topology of non-Hermitian band structures in two and three dimensions}
\author{Charles C. Wojcik$^1$}
\author{Kai Wang$^1$}
\author{Avik Dutt$^1$}
\author{Janet Zhong$^2$}
\author{Shanhui Fan$^{1,2}$}\email[Corresponding author: ]{shanhui@stanford.edu}
\
\affiliation{$^{1}$Department of Electrical Engineering, Ginzton Laboratory, Stanford University, Stanford, CA 94305, USA}
\affiliation{$^{2}$Department of Applied Physics, Stanford University, Stanford, CA 94305, USA}
\date{\today}
 
\begin{abstract}
In the band theory for non-Hermitian systems, the energy eigenvalues, which are complex, can exhibit non-trivial topology which is not present in Hermitian systems. In one dimension, it was recently noted theoretically and demonstrated experimentally that the eigenvalue topology is classified by the braid group. The classification of eigenvalue topology in higher dimensions, however, remained an open question. Here, we give a complete description of eigenvalue topology in two and three dimensional systems, including the gapped and gapless cases. We reduce the topological classification problem to a purely computational problem in algebraic topology. In two dimensions, the Brillouin zone torus is punctured by exceptional points, and each nontrivial loop in the punctured torus acquires a braid group invariant. These braids satisfy the constraint that the composite of the braids around the exceptional points is equal to the commutator of the braids on the fundamental cycles of the torus. In three dimensions, there are exceptional knots and links, and the classification depends on how they are embedded in the Brillouin zone three-torus. When the exceptional link is contained in a contractible ball, the classification can be expressed in terms of the knot group of the link. Our results provide a comprehensive understanding of non-Hermitian eigenvalue topology in higher dimensional systems, and should be important for the further explorations of topologically robust open quantum and classical systems.
\end{abstract}

\maketitle

\begin{figure}[t]
    \centering
    \sidesubfloat[\label{fig:fig1a}]{
        \includegraphics[width=0.41\textwidth]{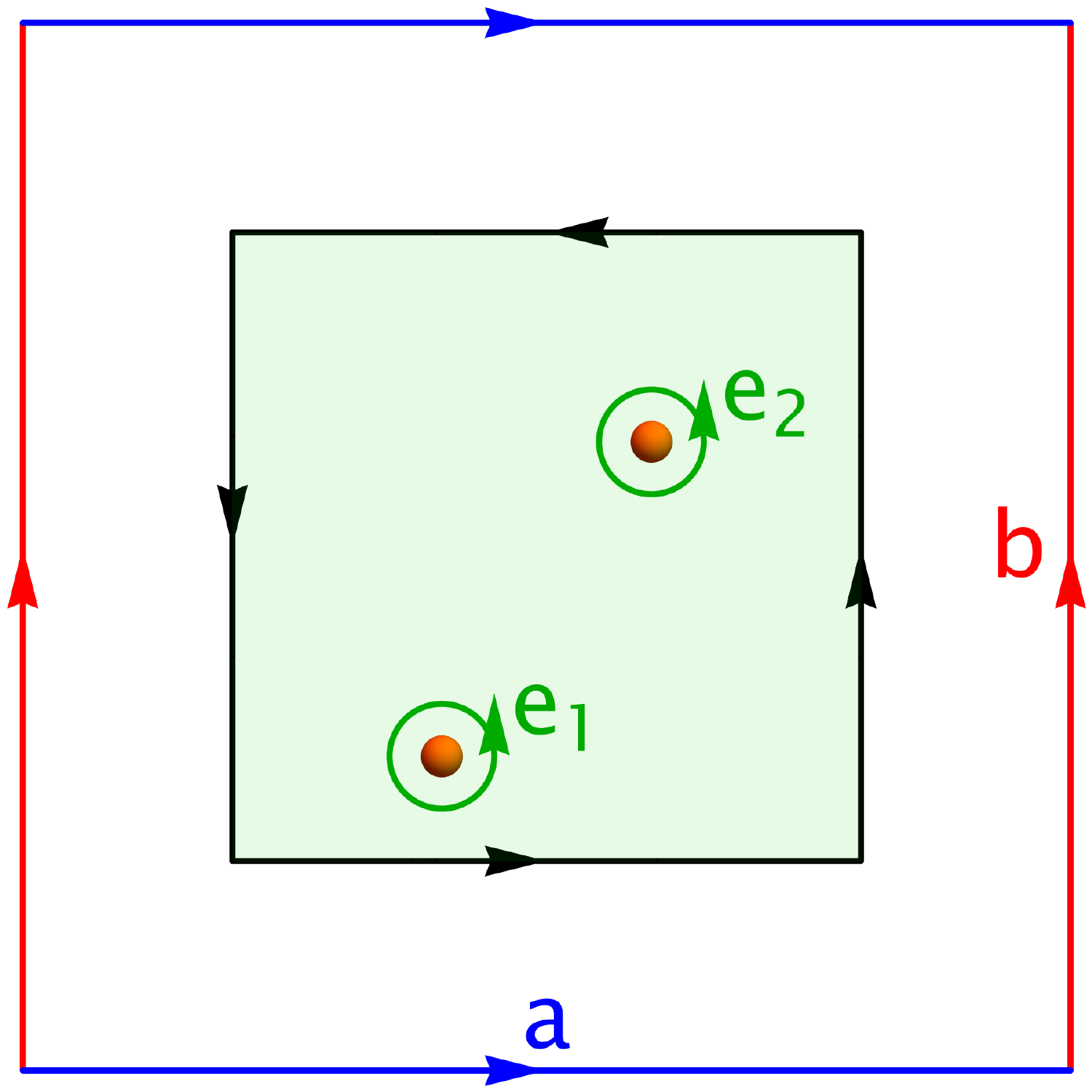}
    }\hfill
    \sidesubfloat[\label{fig:fig1b}]{
    \hspace*{-2em}
    \includegraphics{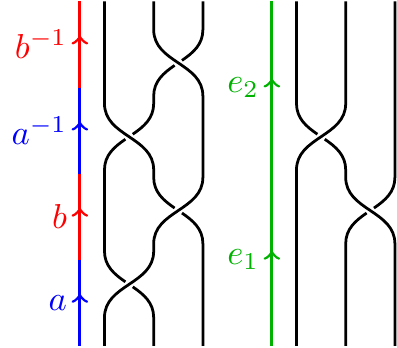}
    }\hfill
    
    \caption{Eigenvalue topology in two dimensions. (a) The Brillouin zone torus is shown as its fundamental polygon. In this example, there are two exceptional points. The nontrivial loops in the twice-punctured torus are the fundamental cycles $a$ and $b$, as well as the loops $e_i$ around the exceptional points. These loops satisfy the relation $aba^{-1}b^{-1} = e_1 e_2$. (b) The eigenvalues of the Hamiltonian define an element of the braid group for each nontrivial loop in the twice-punctured torus. In this three-band example, $a = \sigma_1$ and $b = \sigma_2$ are positive half twists; on the other hand, $e_1 = \sigma_2^{-1}$ and $e_2 = \sigma_1$. We can verify that $\sigma_1 \sigma_2 \sigma_1^{-1} \sigma_2^{-1} = \sigma_2^{-1} \sigma_1$ using the braid relation $\sigma_2 \sigma_1 \sigma_2 = \sigma_1 \sigma_2 \sigma_1$: $\sigma_1 \sigma_2 \sigma_1^{-1} \sigma_2^{-1} = \sigma_2^{-1} (\sigma_2 \sigma_1 \sigma_2) \sigma_1^{-1} \sigma_2^{-1} =  \sigma_2^{-1} (\sigma_1 \sigma_2 \sigma_1) \sigma_1^{-1} \sigma_2^{-1} = \sigma_2^{-1} \sigma_1$. This can also be seen intuitively by "pulling" the ends of the two composite braids.}
    \label{fig:fig1}
\end{figure}

Topological invariants associated with band structures are widely used to predict robust behavior in electronic and photonic systems~\cite{hasan2010colloquium, qi2011topological, lu2014topological, ozawa2019topological, bansil2016colloquium}. For example, in the two-dimensional Chern insulator, the topological invariants of the bulk bands guarantee the existence of chiral edge states~\cite{thouless1982quantized}. In the three-dimensional Weyl semimetal, the topological charges of the Weyl points lead to surface Fermi arcs~\cite{wan2011topological}. These concepts are traditionally studied in Hermitian systems (closed systems without gain or loss). However, there has been much recent interest in generalizing topological band theory to non-Hermitian systems~\cite{gong2018topological, shen2018topological, borgnia2020non, kawabata2019symmetry, okuma2020topological, yao2018edge, longhi2019probing}, which are of particular interest in areas such as photonics, electrical circuits, and mechanics where gain and loss are ubiquitous~\cite{bahari2017nonreciprocal, feng2017non, el2019dawn, harari2018topological, bandres2018topological, zhao2019non, budich2020non, ezawa2019non, ghatak2020observation}.

Non-Hermitian systems have complex eigenvalues and therefore can exhibit nontrivial eigenvalue topology: the eigenvalues can wind and braid around each other in the complex energy plane~\cite{sun2020alice, wojcik2020homotopy, yin2018geometrical, hu2021knots}. This stands in contrast to Hermitian systems, where the topological invariants are defined only in terms of eigenvectors. Non-Hermitian eigenvalue topology has been observed experimentally~\cite{ghatak2020observation, wang2021generating, wang2021topological}. Eigenvalue winding and braiding has also been shown to effect the open boundary condition spectrum resulting in the non-Hermitian skin effect~\cite{yao2018edge, okuma2020topological, weidemann2020topological, zhong2021nontrivial}.

In one spatial dimension, it was recently shown that eigenvalue topology is classified by the braid group~\cite{wojcik2020homotopy}. The nature of eigenvalue topology in two and three dimensions has not yet been established, although simpler point-gap winding number invariants have been explored~\cite{shnerb1998winding, zhang2020correspondence, zhang2021universal, sun2021geometric}. Certain features are known which suggest that two and three dimensions will exhibit richer eigenvalue topology than one dimension. For example, in two dimensions, an exceptional point has a nontrivial topological charge associated with eigenvalue braiding on a small circle enclosing it~\cite{leykam2017edge}. In three dimensions, the exceptional locus is generically one-dimensional, for example forming exceptional rings, lines, or even knots and links~\cite{cerjan2018effects, carlstrom2018exceptional, carlstrom2019knotted, zhang2021tidal}. In addition to these topologically charged exceptional objects, braid-group valued invariants can be associated to non-contractible cycles in the Brillouin zone torus~\cite{wojcik2020homotopy}. Despite this rich behavior and theoretical interest, there is not a comprehensive understanding of eigenvalue topology in two and three dimensions.

In this Letter, we give a complete description of the eigenvalue topology of two and three dimensional band structures, treating both the gapped and gapless settings in a unified way and not assuming any symmetry. We show that the problem of classifying eigenvalue topology of non-Hermitian band structures can be reduced to a computational problem in algebraic topology involving fundamental groups. We provide several examples to illustrate this general scheme.

Our strategy is to isolate the exceptional locus $\Delta$, defined as the set of wavevectors in the Brillouin zone torus $\mathbb{T}^d$ for which there is an eigenvalue degeneracy, and study the eigenvalue topology on the complement $\mathbb{T}^d - \Delta$. The exceptional locus generically has codimension $2$ since it is defined by a single complex equation, the vanishing of the discriminant~\cite{wojcik2020homotopy}. On $\mathbb{T}^d - \Delta$, there is no eigenvalue degeneracy, so the distinct eigenvalues define a point in the unordered configuration space $\UConf_N(\mathbb{C}) = \{\{E_1, \ldots, E_N\} : E_i \in \mathbb{C}, E_i \neq E_j\}$, where $N$ is the number of energy bands~\cite{wojcik2020homotopy}. The fundamental group of this space is the braid group $B_N$ ($\pi_1(\UConf_N(\mathbb{C})) = B_N$), and all higher homotopy groups vanish~\cite{kassel2008braid}. In general, a connected pointed space $Y$ having $\pi_1(Y) = G$ and $\pi_i(Y) = 0$ for $i > 1$ is called an Eilenberg-MacLane space of type $K(G, 1)$, and one writes $Y = K(G, 1)$. Thus
\begin{align}
\UConf_N(\mathbb{C}) = K(B_N, 1).
\end{align}
An important property of the spaces $K(G, 1)$ is that for any path-connected pointed space $X$ there is a natural bijection $[X, K(G, 1)] = \Hom(\pi_1(X), G)$ between the set $[X, K(G, 1)]$ of homotopy classes of maps $X \to K(G, 1)$ and the set $\Hom(\pi_1(X), G)$ of group homomorphisms $\pi_1(X) \to G$~\cite[p.~427, Thm.~9]{spanier1989algebraic}. Applying this result in our setting yields
\begin{align}\label{eq:eq1}
[\mathbb{T}^d - \Delta, \UConf_N(\mathbb{C})] = \Hom(\pi_1(\mathbb{T}^d - \Delta), B_N).
\end{align}
%This formula reduces the topological classification problem to the purely algebraic problem of computing homomorphisms to the braid group. 
As a technical point, the maps in Eq.~\ref{eq:eq1} are assumed to be basepoint-preserving; in order to relax this condition, one must identify any two homomorphisms which are similar under the action of $B_N$ by conjugation.

Eq.~\ref{eq:eq1} is our primary mathematical result: the left-hand side is the set of homotopy classes of non-Hermitian band structures with fixed exceptional locus, which is what we want to know, and the right-hand side is something we are able to calculate. To calculate the right-hand side, the first step is to write down a presentation of the fundamental group $\pi_1(\mathbb{T}^d - \Delta)$ in terms of generators and relations; if $\Delta$ is relatively complicated, the Seifert-Van Kampen theorem may be used here~\cite{hatcher2001algebraic}. Generators of $\pi_1(\mathbb{T}^d - \Delta)$ correspond to specific loops around $\Delta$ in the Brillouin zone torus, and relations correspond to homotopies of composites of these loops. The second step is to compute the set of group homomorphisms $\Hom(\pi_1(\mathbb{T}^d - \Delta), B_N)$. The defining property of a group homomorphism is that $f(\gamma_1 \gamma_2) = f(\gamma_1) f(\gamma_2)$. Because of this property, any group homomorphism $f : \pi_1(\mathbb{T}^d - \Delta) \to B_N$ is determined by its values on the generators of $\pi_1(\mathbb{T}^d - \Delta)$. Conversely, any assignment of braids to the generators of $\pi_1(\mathbb{T}^d - \Delta)$ which satisfies the relations of $\pi_1(\mathbb{T}^d - \Delta)$ gives rise to such a group homomorphism. Thus, in general, elements of $\Hom(\pi_1(\mathbb{T}^d - \Delta), B_N)$ are given by assigning braids to some specific (irreducible) loops around $\Delta$ in the Brillouin zone torus, in such a way that whenever any two composites formed from these generating loops are homotopic, the corresponding braid products are equal.

Fig.~\ref{fig:fig1} shows the case of two dimensions. The Brillouin zone is a torus $\mathbb{T}^2$, and the exceptional locus $\Delta$ generically consists of a finite number of exceptional points. (Other configurations are possible by fine-tuning, but they are non-generic in the sense that they are removed by arbitrarily small perturbations). If $k$ is the number of exceptional points, then $\mathbb{T}^2 - \Delta$ is a $k$-punctured torus. We first consider the fully-gapped case $k = 0$. The fundamental group $\pi_1(\mathbb{T}^2) = \mathbb{Z}^2$ is the free abelian group generated by the two fundamental cycles $\gamma_a$ and $\gamma_b$: its elements have the form $\gamma_a^m \gamma_b^n$ for some integers $m$ and $n$. The reason that this fundamental group is abelian is that there is a two-cell connecting the curve $\gamma_a \gamma_b$ to the curve $\gamma_b\gamma_a$, giving rise to the relation $\gamma_a\gamma_b = \gamma_b\gamma_a$. If $f : \pi_1(\mathbb{T}^2) \to B_N$ is a group homomorphism, then $f(\gamma_a)$ and $f(\gamma_b)$ commute, since $f(\gamma_a)f(\gamma_b) = f(\gamma_a \gamma_b) = f(\gamma_b \gamma_a) = f(\gamma_b) f(\gamma_a)$; so we get a pair of commuting braids. Conversely, for any pair of commuting braids $a$ and $b$, we can define a group homomorphism $f : \pi_1(\mathbb{T}^2) \to B_N$ by setting $f(\gamma_a) = a$ and $f(\gamma_b) = b$; then $f(\gamma_a^m \gamma_b^n) = a^m b^n$. We can express the fact that $a$ and $b$ commute concisely in terms of the commutator $[a, b] = aba^{-1}b^{-1}$; they commute precisely when $[a, b] = 1$. It follows that the eigenvalue topology in the fully gapped case is given by commuting pairs of braids:
\begin{align}\label{eq:eq3}
\Hom(\pi_1(\mathbb{T}^2), B_N) = \{a, b \in B_N : [a, b] = 1\}.
\end{align}
Eq.~\ref{eq:eq3} shows the richness of eigenvalue topology even in the simple case of gapped two-dimensional systems. While there are certain pairs of braids which obviously commute, such as those that act entirely on different strands, these are not the only ones; due to the complexity of the braid group, determining pairs of commuting braids is already an interesting mathematical problem~\cite{franco2003computation}.

\begin{figure}[t]
    \centering
    \sidesubfloat[\label{fig:fig2a}]{
        \includegraphics[width=0.41\textwidth]{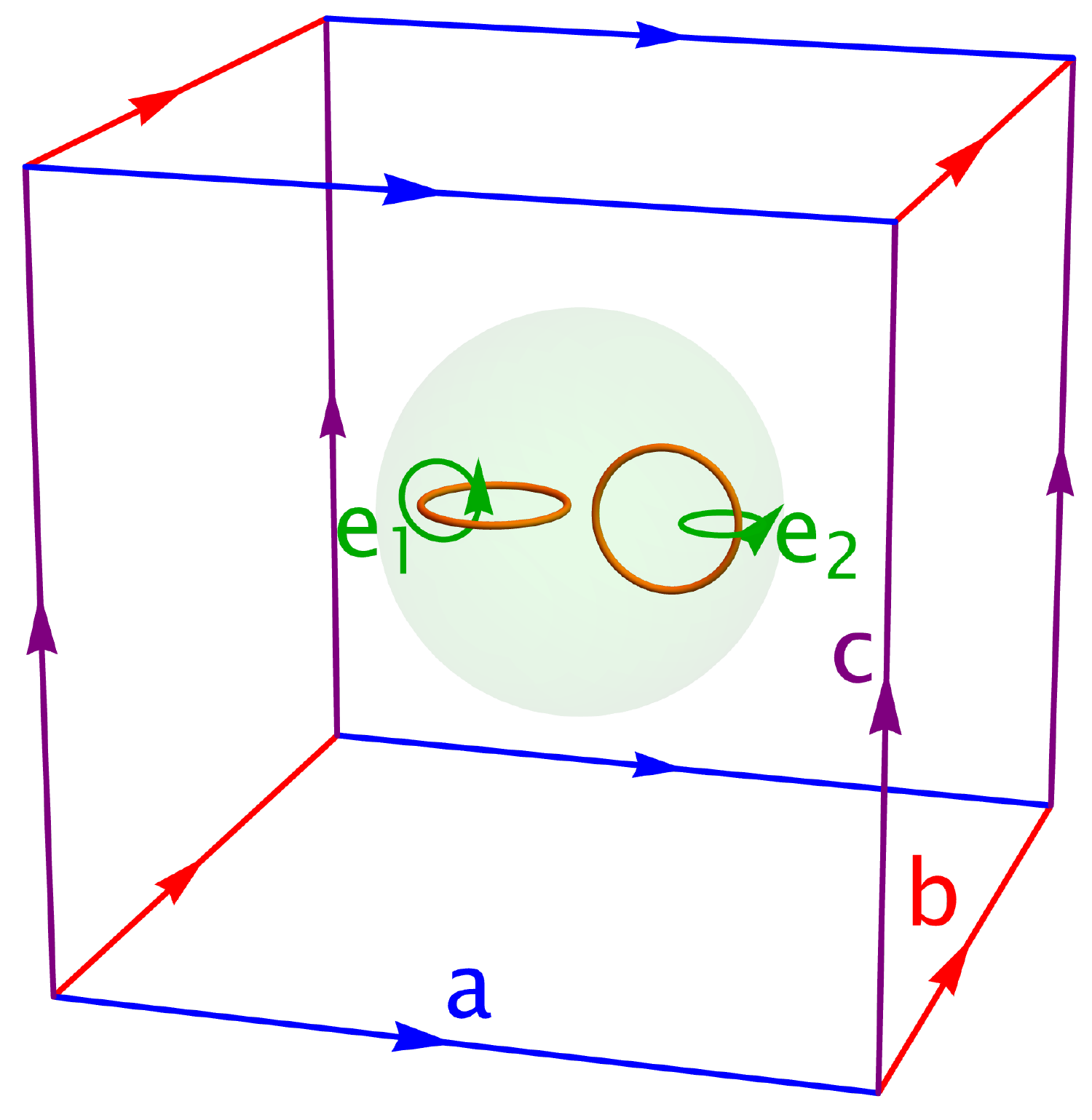}
    }\hfill
    \sidesubfloat[\label{fig:fig2b}]{
        \includegraphics[width=0.41\textwidth]{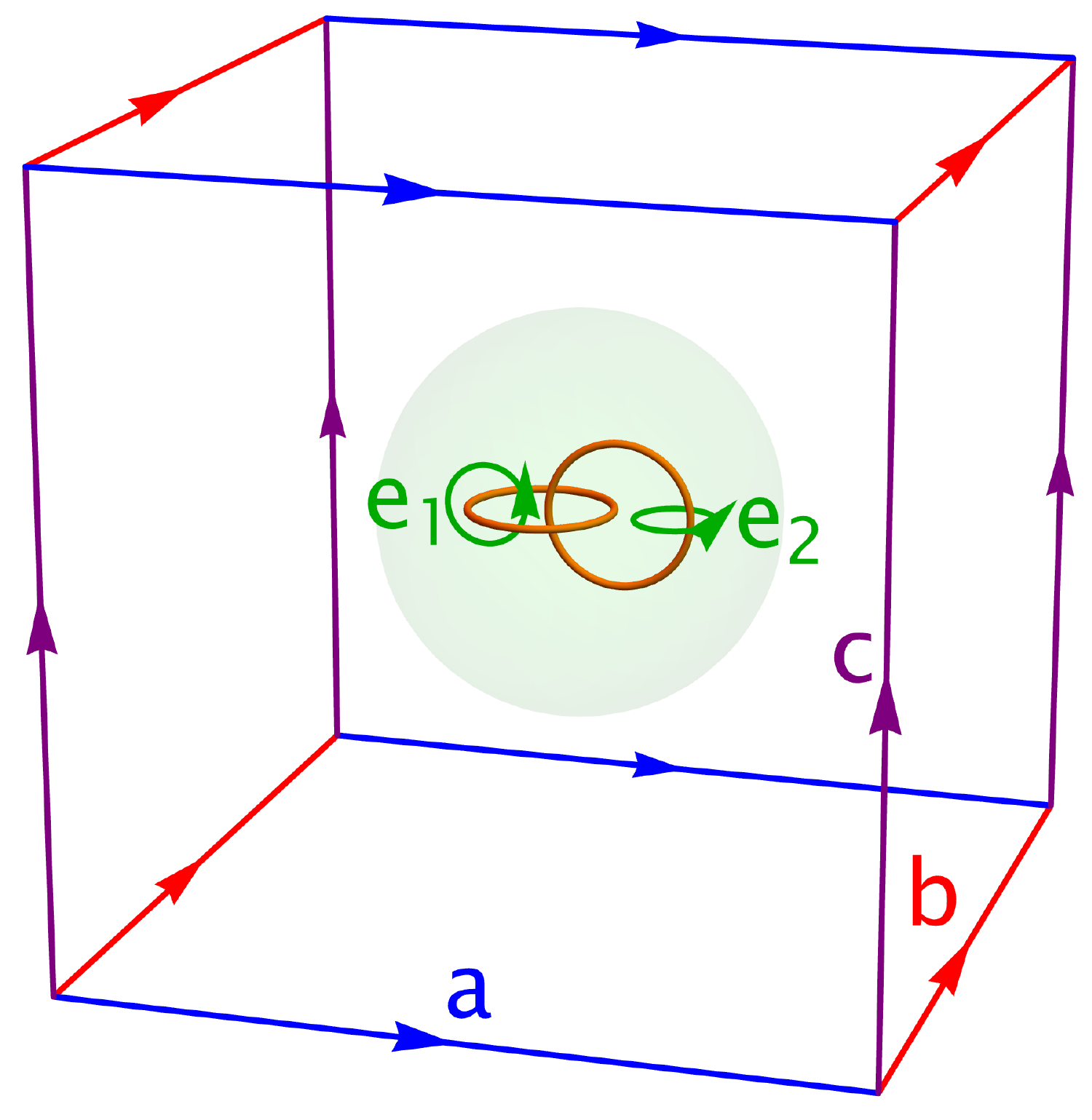}
    }\hfill
    \sidesubfloat[\label{fig:fig2c}]{
        \includegraphics[width=0.41\textwidth]{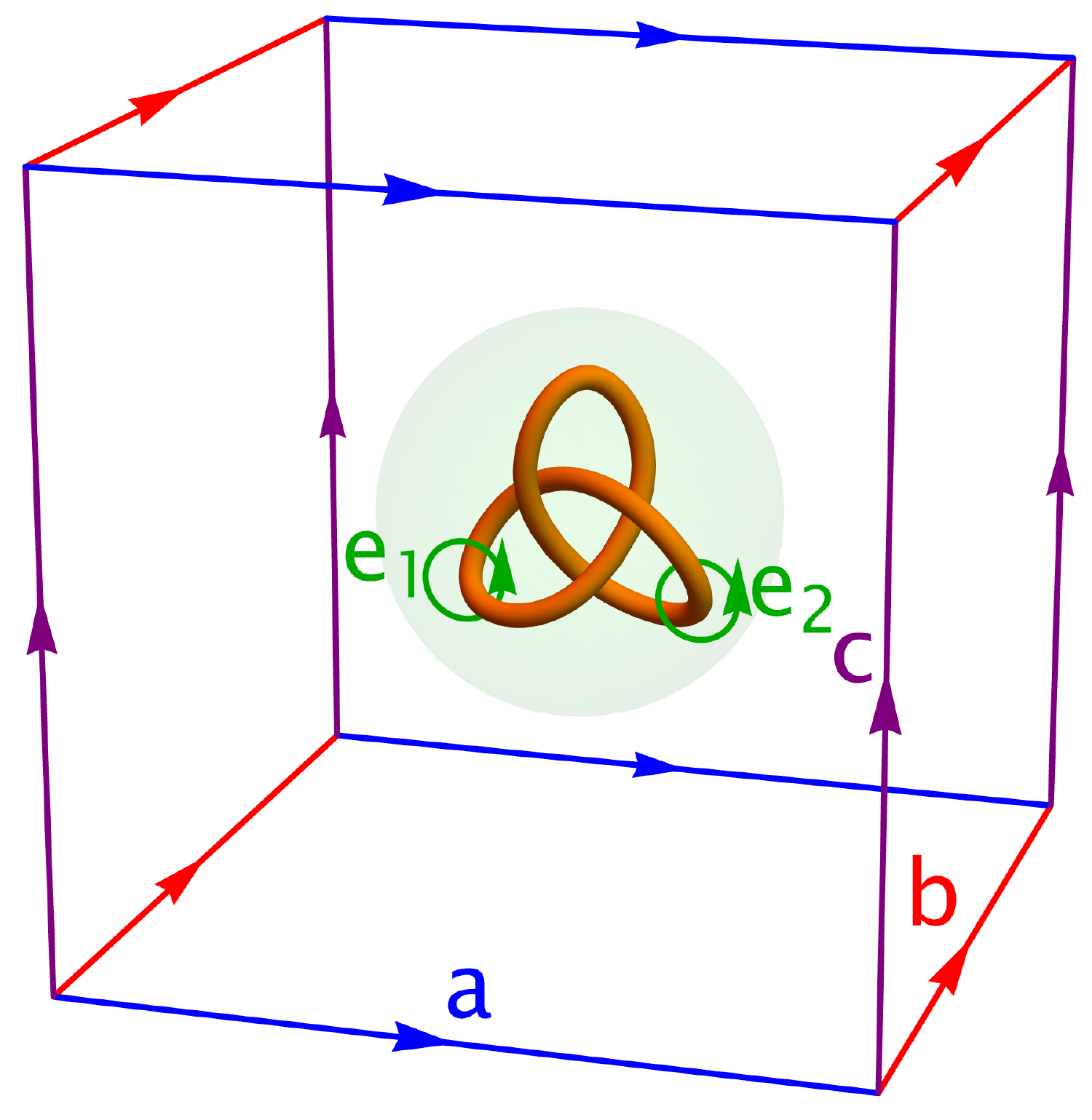}
    }\hfill
    \sidesubfloat[\label{fig:fig2d}]{
        \includegraphics[width=0.41\textwidth]{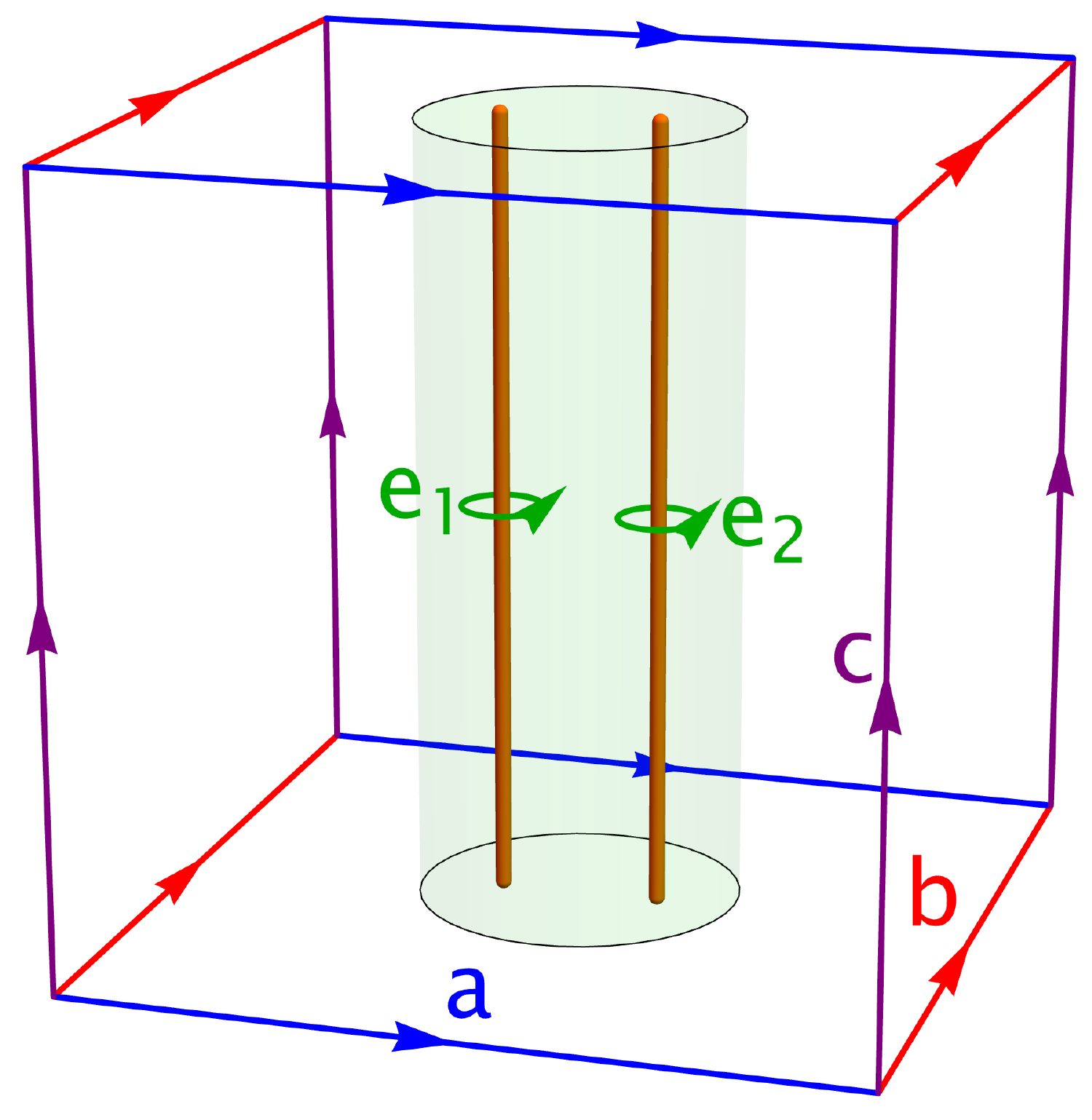}
    }\hfill
%\begin{figure}[t]
%    \centering
%    \includegraphics[width=0.99\textwidth]{fig2.png}
    \caption{Eigenvalue topology in three dimensions. Four different exceptional links are shown in panels (a)-(d): the unlink, the Hopf link, the trefoil knot, and a pair of nontrivial cycles. In the first three cases (a)-(c), the exceptional link $\Delta$ is contained in a small ball; the nontrivial loops are the fundamental cycles $a, b,$ and $c$, as well as the generators $e_i$ of the knot group $\pi_1(\mathbb{R}^3 - \Delta)$. The relations are $[a,b] = [b,c] = [c,a] = 1$, together with any relations coming from the knot group. In the final case (d), the exceptional link is contained in a solid torus rather than a ball, leading to the relations $[a, b] = e_1 e_2$, $[b,c] = [c,a] = 1$.}
    \label{fig:fig2}
\end{figure}

Now we consider the gapless case $k \geq 1$. In this case, each exceptional point acts like a puncture in the torus; the complement of the punctures can be deformation retracted onto a one-dimensional skeleton which is simply $k+1$ circles joined at a point~\cite{hatcher2001algebraic}. The fundamental group of this $k$-times punctured torus is $\pi_1(\mathbb{T}^2 - \Delta) = \mathbb{Z}^{*{(k+1)}}$, the free group on $k+1$ generators. In Fig.~\ref{fig:fig1}(a), we show the obvious cycles $\gamma_{e_1}, \ldots, \gamma_{e_k}$ around each exceptional point in addition to the cycles $\gamma_a$ and $\gamma_b$. Each cycle defines an element of $\mathbb{Z}^{*{(k+1)}}$, but since there are $k+2$ cycles, they are not all independent. Indeed, the cycles satisfy the relation $[\gamma_a, \gamma_b] = \gamma_{e_1} \cdots \gamma_{e_k}$. This relation arises from the fact that both $[\gamma_a, \gamma_b]$ and $\gamma_{e_1} \cdots \gamma_{e_k}$ can be deformed to the boundary $\partial U$, where $U$ is the green region in Fig.~\ref{fig:fig1}(a). Now we compute group homomorphisms to the braid group: as before, the generators of $\pi_1(\mathbb{T}^2 - \Delta)$ are assigned to braids subject to constraints coming from the relations. The eigenvalue topology in the gapless case is therefore
\begin{align}\label{eq:eqgapless}
& \Hom(\pi_1(\mathbb{T}^2 - \Delta), B_N) =\\
& \{a, b, e_1, \ldots, e_k \in B_N : [a, b] = e_1 \cdots e_k\}. \nonumber
\end{align}
This can also be written as $\Hom(\pi_1(\mathbb{T}^2 - \Delta), B_N) = \{a, b, e_1, \ldots, e_{k-1} \in B_N\}$ with the understanding that $e_k = (e_1 \cdots e_{k-1})^{-1} [a, b]$ in order to satisfy the constraint from Eq.~\ref{eq:eqgapless}. This is what one would obtain directly from the formula $\pi_1(\mathbb{T}^2 - \Delta) = \mathbb{Z}^{*{(k+1)}}$. Specializing Eq.~\ref{eq:eqgapless} to the two-band case $N = 2$, where $B_2 = \mathbb{Z}$, we obtain
\begin{align}
&\Hom(\pi_1(\mathbb{T}^2 - \Delta), B_2) = \\
& \{a, b, e_1, \ldots, e_k \in \mathbb{Z} : 0 = \sum_{i=1}^k e_i\}. \nonumber
\end{align}

We note that Eq.~\ref{eq:eqgapless} includes the gapped and gapless cases. We also note that this formula implies the doubling theorem for exceptional points~\cite{yang2021fermion}: the product $e_1 \cdots e_k$ is a commutator so it has total degree zero, but each $e_i$ has degree $\pm 1$ (since generically it is conjugate to a half twist), so the exceptional points come in pairs. (The reason each $e_i$ is conjugate to a generator and not necessarily itself a generator is a technical one: all loops must be connected to the basepoint in order to have a group structure, so the braid on the loop is conjugated by the braid on the path to the basepoint.) %The converse is not true: not every braid with vanishing total degree has commutator length one~\cite{calegari2008stable, brandenbursky2015concordance}. 

The situation in three dimensions is far richer, as shown in Fig.~\ref{fig:fig2}. The Brillouin zone is now a 3-torus $\mathbb{T}^3$, and the exceptional locus $\Delta$ is generically a one-dimensional submanifold, namely an exceptional knot or more generally an exceptional link. The gapped case is a straightforward generalization from two dimensions: $\pi_1(\mathbb{T}^3) = \mathbb{Z}^3$, so
\begin{align}
&\Hom(\pi_1(\mathbb{T}^3), B_N) = \\
& \{a, b, c \in B_N : [a, b] = [b, c] = [c, a] = 1\}. \nonumber
\end{align}

The gapless case in three dimensions is more complicated in general. To illustrate the general structure of the solutions we first consider a few examples. Suppose that the exceptional link $\Delta$ is contained in a small ball $U$. In this case, we can compute $\pi_1(\mathbb{T}^3 - \Delta)$ using the Seifert-Van Kampen theorem from algebraic topology~\cite{hatcher2001algebraic}. The theorem allows us to compute $\pi_1(U_1 \cup U_2)$ for open sets $U_1$ and $U_2$ in terms of $\pi_1(U_1)$, $\pi_1(U_2)$, and $\pi_1(U_1 \cap U_2)$. In our setting, we take $U_2 = U - \Delta$, and we take $U_1$ to be a thickening of $\mathbb{T}^3 - U$. Then $U_1 \cup U_2 = \mathbb{T}^3 - \Delta$, and $U_1 \cap U_2$ deformation retracts onto the boundary $\partial U$. Since $U$ is a ball, its boundary is a sphere, which is simply connected: $\pi_1(U_1 \cap U_2) = 0$. Because of this, the Seifert-Van Kampen theorem tells us that $\pi_1(U_1 \cup U_2)$ is simply a free product $\pi_1(U_1) * \pi_1(U_2)$, meaning that loops in $U_1$ and $U_2$ can be combined with no additional relations. Furthermore, since $U - \Delta$ is homeomorphic to the complement of a link in $\mathbb{R}^3$, we have $\pi_1(U - \Delta) = \pi_1(\mathbb{R}^3 - \Delta)$, the knot group of the link $\Delta$~\cite{stillwell2012classical}. Finally, we note that removing $U$ from $\mathbb{T}^3$ has no effect on the fundamental group, namely $\pi_1(\mathbb{T}^3 - U) = \pi_1(\mathbb{T}^3) = \mathbb{Z}^3$. Putting this together, we have
\begin{align}
\pi_1(\mathbb{T}^3 - \Delta) = \mathbb{Z}^3 * \pi_1(\mathbb{R}^3 - \Delta).
\end{align}
Now we compute homomorphisms to the braid group. The free product means that there are no relations between the braids assigned to the knot group $\pi_1(\mathbb{R}^3 - \Delta)$ and the braids assigned to the torus cycles $\mathbb{Z}^3$, namely
\begin{align}\label{eq:eqball}
&\Hom(\pi_1(\mathbb{T}^3 - \Delta), B_N) = \\
&\{a, b, c \in B_N, f \in \Hom(\pi_1(\mathbb{R}^3 - \Delta), B_N): \nonumber \\
&[a, b] = [b, c] = [c, a] = 1\}.  \nonumber
%&\Hom(\pi_1(\mathbb{T}^3), B_N) \times \Hom(\pi_1(\mathbb{R}^3 - \Delta), B_N). \nonumber
\end{align}

Fig.~\ref{fig:fig2}(a)-(c) show three examples of an exceptional link that can be surrounded by a contractible ball $U$. For the unlink (Fig.~\ref{fig:fig2}(a)), we have $\pi_1(\mathbb{R}^3 - \Delta) = \mathbb{Z}^{*2}$, a free group on loops around the two components, so 
\begin{align}
&\Hom(\pi_1(\mathbb{T}^3 - \Delta), B_N) = \\
&\{a, b, c, e_1, e_2 \in B_N : [a, b] = [b, c] = [c, a] = 1\}. \nonumber
\end{align}
For the Hopf link (Fig.~\ref{fig:fig2}(b)), the crossing gives rise to the relation $e_1 e_2 = e_2 e_1$, so $\pi_1(\mathbb{R}^3 - \Delta) = \mathbb{Z}^2$ is a free abelian group and 
\begin{align}
&\Hom(\pi_1(\mathbb{T}^3 - \Delta), B_N) = \\
&\{a, b, c, e_1, e_2 \in B_N : \nonumber \\
&[a, b] = [b, c] = [c, a] = 1, [e_1, e_2] = 1\}. \nonumber
\end{align}
Finally, for the trefoil knot (Fig.~\ref{fig:fig2}(c)), the knot group is itself the braid group on three strands, $\pi_1(\mathbb{R}^3 - \Delta) = B_3$. This group is generated by $\sigma_1$ and $\sigma_2$ with the braid relation $\sigma_1 \sigma_2 \sigma_1 = \sigma_2 \sigma_1 \sigma_2$, so 
\begin{align}
&\Hom(\pi_1(\mathbb{T}^3 - \Delta), B_N) = \\
&\{a, b, c, e_1, e_2 \in B_N : \nonumber \\
&[a, b] = [b, c] = [c, a] = 1, e_1 e_2 e_1 = e_2 e_1 e_2\}. \nonumber
\end{align}

Other possibilities for $\Delta$ can also be tractable; for example, suppose $\Delta$ itself is a pair of nontrivial cycles, as shown in Fig.~\ref{fig:fig2}(d). In this case, we can use the Seifert-Van Kampen theorem as before, but with $U$ a solid torus, so that $\pi_1(\partial U) = \mathbb{Z} \times \mathbb{Z}$. In this case, $\pi_1(\mathbb{T}^3 - \Delta)$ is not a free product but rather an amalgamated product~\cite{hatcher2001algebraic}; one of the $\mathbb{Z}$ factors of $\pi_1(\partial U)$ is trivially identified with $\gamma_c$, and the other leads to the relation $[\gamma_a, \gamma_b] = \gamma_{e_1} \gamma_{e_2}$. It follows that
\begin{align}\label{eq:eqtorus}
&\Hom(\pi_1(\mathbb{T}^3 - \Delta), B_N) = \\
&\{a, b, c, e_1, e_2 \in B_N : [a, b] = e_1 e_2, [b, c] = [c, a] = 1\}. \nonumber
\end{align}
We note that $\Delta$ cannot be a single nontrivial cycle since any two-torus section must obey the doubling theorem~\cite{yang2021fermion} (exceptional points must come in pairs). On the other hand, a single exceptional ring or a single trefoil knot is allowed from the perspective of eigenvalue topology; but these configurations still may be forbidden due to constraints from eigenvector topology if they have nonzero Chern number~\cite{bansil2016colloquium}.

We can generalize the previous discussion to apply the Seifert-Van Kampen theorem to an arbitrary exceptional link contained in an open set $U$, possibly in a higher-dimensional Brillouin zone. For general $U$ using $U_1$ and $U_2$ as defined above, the Seifert-Van Kampen theorem says that
\begin{align}\label{eq:eqsvk}
\pi_1(\mathbb{T}^d - \Delta) = \pi_1(\mathbb{T}^d - U) *_{\pi_1(\partial U)} \pi_1(U - \Delta)
\end{align}
where $G *_K H$ denotes the pushout of groups $G$ and $H$ over a group $K$~\cite{hatcher2001algebraic}. By the universal property of the pushout~\cite{lang02}, 
\begin{align}\label{eq:eqpullback}
&\Hom(\pi_1(\mathbb{T}^d - \Delta), B_N) = \widetilde{G} \times_{\widetilde{K}} \widetilde{H} 
%&=\{(\widetilde{g}, \widetilde{h}) \in \widetilde{G} \times \widetilde{H} : \nonumber \\
%&\widetilde{g} \textrm{ and } \widetilde{h} \textrm{ agree when mapped to } \widetilde{K}\} \nonumber
\end{align}
where
\begin{align}
\widetilde{G} &= \Hom(\pi_1(\mathbb{T}^d - U), B_N) \\
\widetilde{H} &= \Hom(\pi_1(U - \Delta), B_N) \nonumber \\
\widetilde{K} &= \Hom(\pi_1(\partial U), B_N). \nonumber
\end{align}
In Eq.~\ref{eq:eqpullback}, $\widetilde{G} \times_{\widetilde{K}} \widetilde{H}$ denotes the pullback (fiber product) of groups $\widetilde{G}$ and $\widetilde{H}$ over a group $\widetilde{K}$~\cite{lang02}. This pullback is the subset of the Cartesian product $\widetilde{G} \times \widetilde{H}$ consisting of pairs which are compatible when passed to $\widetilde{K}$. In this way, when we apply the Seifert-Van Kampen theorem to understand more complicated systems, the results agree with our general notion that braids should be assigned to irreducible non-contractible cycles in $\mathbb{T}^d - \Delta$ subject to relations coming from homotopies between composite cycles.

There is a rich set of effects of eigenvalue topology on the physics of non-Hermitian systems. The spectrum of a finite-size non-Hermitian system looks quite different from the spectrum of an infinite system due to the non-Hermitian skin effect~\cite{okuma2020topological, yao2018edge}. In one dimension, eigenvalue braiding has been observed to control certain aspects of the shape of the finite-size spectrum as a subset of the complex energy plane~\cite{zhong2021nontrivial, wang2021topological}. Furthermore, eigenvalue topology may modify the bulk-edge correspondence for eigenvector topology, since the eigenvalues determine the generalized Brillouin zone~\cite{yokomizo2019non, yang2020non}. Finally, there is a rich interplay between eigenvalue topology and eigenvector topology, resulting in a reduction mod $N$ of the Chern number in the presence of braiding~\cite{sun2020alice, wojcik2020homotopy}. Understanding eigenvalue topology in two and three dimensions is therefore key for understanding non-Hermitian topology more broadly.

In summary, we have provided a complete description of non-Hermitian eigenvalue topology in two and three dimensions in both gapped and gapless systems. The topological classification problem reduces to a well-studied problem in algebraic topology. We provided concrete examples in two and three dimensions to illustrate the general scheme. These results should be helpful in developing a more general understanding of non-Hermitian topology.

This work is supported by the Vannevar Bush Faculty Fellowship from the U. S. Department of Defense (Grant No. N00014-17-1-3030), and by a Simons Invesigator in Physics grant from the Simons Foundation (Award No. 827065).

\bibliography{bib}{}
\end{document}